\documentclass{ws-ijmpd}
\usepackage[super,compress]{cite}
\usepackage{subfigure}
\usepackage{url}
\usepackage{graphicx}
\usepackage{float}
\usepackage{hyperref}
\UseRawInputEncoding 

\def\be{\begin{equation}}
	\def\ee{\end{equation}}
\def\bea{\begin{eqnarray}}
	\def\eea{\end{eqnarray}}

\begin{document}


%
%

\title{\boldmath Internal heating mechanisms in neutron stars  }

\author{F. K\"opp}

\address{Instituto de F\'isica, Universidade Federal do Rio Grande do Sul,
	Av. Bento Gon\c{c}alves 9500, 91501-970,
	Porto Alegre, RS, Brazil \\ {\it{fabiokopp@proton.me}}
}

\author{J.E. Horvath}

\address{Instituto de Astronomia, Geof\'isica e Ci\^encias Atmosf\'ericas,
	CEP 05508-900, S\~ao Paulo, SP, Brazil \\  {\it{foton@astro.iag.usp.br
	}}
}

\author{D. Hadjimichef}
\address{Instituto de F\'isica, Universidade Federal do Rio Grande do Sul,
	Av. Bento Gon\c{c}alves 9500, 91501-970,
	Porto Alegre, RS, Brazil \\ {\it{dimiter.hadjimichef@ufrgs.br}}
}

\author{C.A.Z. Vasconcellos}
\address{Instituto de  F\'isica, Universidade Federal do Rio Grande do Sul,
	Av. Bento Gon\c{c}̧alves 9500, 91501-970,
	Porto Alegre, RS, Brazil and \\ International Center for Relativistic
	Astrophysics Network (ICRANet),
	Pescara, Italy
	 \\ {\it{cesarzen@cesarzen.com}}
}
\author{P. O. Hess}
\address{Instituto de Ciencias Nucleares, Universidad Nacional Aut́\'anoma de M\'exico,
	Circuito Exterior, C.U., A.P. 70-543, 04510 M\'exico D.F., Mexico  and  \\
	Frankfurt Institute for Advanced Studies, Johann Wolfgang Goethe Universit at,
	Ruth-Moufang-Str. 1, 60438 Frankfurt am Main, Germany \\
	 {\it{hess@nucleares.unam.mx}}
}
\maketitle{}
\begin{history}
	\received{Day Month Year}
	\revised{Day Month Year}
\end{history}

\begin{abstract}
The cooling mechanisms of a neutron star (hereafter referred to as NS) has the potential to reveal important features of superdense matter. The values of the surface temperatures are known for a good sample of NSs aged $\leq 10^6$ years and, with some exceptions, can be explained through standard cooling mechanisms (neutrinos and photons emissions without internal heating), as confirmed in our study. However, for older objects the surface temperatures are in some cases higher than expected, and it is necessary to consider some kind of internal heating to explain these results.With this objective, we revisit in this article the kinetic mechanisms of heating in NSs considering fermionic dark matter heating, rotochemical heating and magnetic field decay. Our results indicate that NSs older than $\sim 10^6$ years, such as some ``Black Widows'' (a subset of binary systems in which only the upper bounds of surface temperatures are known) and old pulsars, in contrast to younger NSs, exhibit much higher surface temperatures than the values predicted by these three heating mechanisms. Furthermore, by restricting the dark matter heating parameters to the current values that were fitted and/or measured for the local dark matter density, masses and neutron star radii, the models studied here also do not reproduce the upper limits of the temperatures from the surface of Black Widows or the actual temperatures of other ancient pulsars. We conclude that if the upper limits for Black Widows are close to real temperatures, dark heating will not represent a convincing explanation of these results, indicating that rotochemical mechanisms may be favored.
\end{abstract}

\ccode{PACS numbers:}

\section{Introduction}
\label{sec:intro}

In the recent decades, the study of compact stars has achieved extraordinary progress as a result of the launching of several technically advanced observational facilities spanning the radio, optical, UV and X-ray regimes, and pushing the limits of our understanding of stellar timing, spectra and imaging. These technological advances and the consequent advances in our knowledge about compact objects have consolidated our understanding of the role of neutron stars and pulsars as unique laboratories for probing matter under extreme conditions of density, gravity, and high-intensity magnetic fields, physical conditions not available in any terrestrial facility.

The death of a massive star in a supernova explosion and the ejection of the stellar envelope results in a neutron star or a black hole, depending on the progenitor mass and other poorly known details\cite{Heger:2002by, NOS}. There may be additional evolution channels not yet fully understood (see ref. \cite{wang}). Once the compact star achieves ``maturity'' (a very short time indeed), the surface of a neutron star  is composed by a  thin atmosphere of hydrogen and helium atoms and possibly of heavier ashes that resulted from the supernova process, resting on a solid crust of heavier atoms. Below this layer, the gravitational pressure is so intense that almost all protons combine with electrons, striped from atomic nuclei, to generate neutrons. In the most internal regions, speculations about its composition have not yet been given a definitive answer: whether neutrons are the dominant particles in an environment composed of protons, neutrons and electrons, or if the extreme gravitational pressure compresses the matter inducing the appearance of more exotic particles like hyperons, the decomposition of protons and neutrons into up, down and strange quarks in almost equal proportions, either existing at high pressure or possibly absolutely stable (strange matter), or even a Bose-Einstein condensate, a state of matter in which all subatomic particles behave as a single quantum mechanical entity.

After decades of speculation, and thanks to a newly launched instrument, the International Space Station Neutron Star Interior Composition Explorer (NICER)~\cite{NICER}, that enables rotation-resolved spectroscopy of the thermal and non-thermal emissions of neutron stars in the soft ($0.2- 12 \, keV$) X-ray band with unprecedented sensitivity,  expectations are growing of probing the interior structure and the origin of dynamic internal phenomena, as well as the composition and global properties of these compact objects. The capabilities the NICER device brings to the investigation of these compact objects are unique: simultaneous fast timing and spectroscopy, with low background and high throughput, also providing more broadly continuity in X-ray timing astrophysics.

Before the advent of NICER, the masses and sizes of neutron stars, essential quantities for discerning different theoretical models of composition, structure and the equation of state (EoS) of these compact objects, were estimated by observing binary star systems using Kepler's laws and light-curve/spectral modeling \cite{jorge}. The determination of masses and radii by the latter method results in considerable uncertainties in  the properties of these objects, and only a few cases (mainly Double Neutron Star systems) could be measured with accuracy. NICER already produced quite precise measurements including radii, allowing an indirect exploration of the exotic states of matter within neutron stars and helping to improve our knowledge about these compact stars. In short, NICER measurements, combined with other multi-messenger observations, can offer significant observational evidence for the quest of the EoS of dense stars and narrow down its microscopic composition.

In the family of binary neutron stars and pulsars, there is a particular group which has received increasing attention in recent years. These are called Black Widow (BW) spiders, a type of system consisting of a millisecond pulsar, in close orbits with a smaller companion star, which is actually being ablated by the pulsar wind. The group of BWs raised a renewed theoretical interest in the study of these systems~\cite{Benvenuto2012, Benvenuto2014, Benvenuto2015, Chen2013, Jia2015, Jia2016, Liu2017, Ablimite2019, Ann2020, Ginzburg2020, Kwan2021, Douglas2022, Kennedy2022, Ray2022, Burdge2022}. Among the characteristics and properties that differentiate Black Widow pulsars, there are hints of high surface temperature (upper limits for now, see below) \cite{Gentile2014}. The standard models of neutron star cooling predict surface temperatures of $\it{T_s}$  $< 10^5$ K for stars ages $\tau$ $> 10^{5-6}$ years~\cite{Potekhin2020, Goldreich, Schaab:1999as}. The confirmed ultraviolet thermal emissions detected from millisecond ``conventional'' pulsars younger than $10^7$ years may imply that BWs do not cool much, in spite of their very old age \cite{Benvenuto2012, Chen2013} if their surface temperature remains around $\it{T_s}$ $\sim 10^5$ K~\cite{Deller2008, Andreas2010}. It appears that Black Widows could feature surface temperatures ranging from one to three orders of magnitude higher~\cite{Gentile2014} than expected, a trend that requires a consistent approach to determine whether the mechanisms responsible for such a phenomenon can explain the measured values, a task that has motivated our present work.

Guided by these observations, in this article we compare the predictions of heating and cooling mechanisms affecting young, mature and Black Widow pulsars to expand our knowledge about these objects under extreme physical conditions.

\section{Dark Matter in galactic halos and neutron stars}

The fundamental nature of dark matter (DM), one of the most fascinating and intricate mysteries of the cosmos, is still unknown. The first evidence for the existence of DM comes from studies involving the movement of galaxy clusters. Zwicky~\cite{Zwicky} found in the 30's that the amount and distribution of luminous matter of the galaxies within the Coma cluster could not provide an explanation for the observed rotation curves. He noticed that the galaxies were moving much faster than their supposedly known masses would logically dictate. Zwick figured there would be additional matter unobservable by conventional methods inside them, some kind of invisible matter that he imaginatively called dark matter. Among others evidences as weak lensing measurements, hot gas in clusters, primordial nucleosynthesis and microwave background experiments, estimates suggest that $\approx 80\%$ of the matter present in the Universe is in the form of DM. Direct DM detection experiments, using some of the most sophisticated equipment, sensitive enough to detect the slightest glimpse of these particles over an increasing range of masses and cross-sections, have so far found no evidence of DM. This has been the main motivation to observe other systems in the Universe, where interactions involving DM could result in observable effects, in other words, in observable signs of these very elusive particles.

As any other star in the galaxy, NSs can attract and accrete DM along their orbital trajectory around the center of the Galaxy. The interaction of DM particles with the Standard Model particles that make up the star, can result in energy losses of the incoming species, trapping and accumulation of them inside NSs due to their intense stellar gravitational field.
The probability of gravitational capture of DM in NSs is further increased due to the very high baryon density of these objects, enhancing this way the interaction capture probability\footnote{The {\it threshold capture cross section} is equivalent to the saturation cross section of the star, i.e.,  $\sigma_{s} =  R^2_{\star} m_N/M_{\star}$, which is of the order $O(10^{-45} cm^2)$~\cite{Bell2018} (where $R_{\star}$ is the radius and $M_{\star}$ the mass of the star; $m_N$ is the nucleon mass).}.
As a result of this accretion process, many DM particles are captured inside the star, and cold NSs may undergo additional internal heating to a detectable level, making these compact objects act at the cosmic scale as ``DM detectors''.

The quest for the existence and nature of DM through its accumulation in NSs is complementary to direct detection processes, since this kind of  search is not restricted by the modeling uncertainties in the nuclear response functions related to the embedding of the DM-nucleon interaction, or uncertainties on the spectral shape of form factors~\cite{Vera2015}, or even by the recoil threshold or the mass of the target~\cite{Bell2018}. Moreover, the ``NS method'' goes beyond direct detection data analysis regarding the momentum dependence of the cross-section, which has been so far mostly focused on simplest scattering scenarios, in which the cross section is momentum-energy and velocity-independent, limiting DM couplings to the nuclear-target through coherent spin-independent (SI) or spin-dependent (SD) interactions. A number of studies have pointed out however that direct detection can access a richer phenomenology, which are manifested through nontrivial momentum or velocity dependence of the scattering cross section and triggering new types of nuclear responses~\cite{Vera2015}. 

The gravitational attraction of NSs in turn accelerates DM particles to relativistic speeds, thus avoiding a linear momentum suppression of the scattering cross-section, and overcoming in this way scenarios of direct detection that emphasize such suppression, guided by scattering rate scales with coherent spin-independent or spin-dependent interactions. On the other hand, this complementarity may be explored, for example, through the investigation of the sensitivity in discriminating SI and SD type processes, which are more effective in direct detection processes~\cite{Bell2018}.

In short, DM in the local galactic halo, accelerated by gravity to over half the speed of light towards a NS may become this way trapped inside by the stellar gravity, accumulating and depositing kinetic energy in the stellar interior, thus heating old, cold neutron stars to higher temperatures, causing this way observable effects of DM onto ordinary matter. The kinetic heating mechanism can even produce some optical emission from NSs in the galactic bulge, and X-ray emission near the galactic center, since the DM density is higher in these regions~\cite{Masha2017}.

The above arguments apply to all NSs, but the case of Black Widows is particularly interesting. As stated, BWs are binary systems composed by a millisecond pulsar (MSP) and a
low-mass companion, with a few per cent of the mass of the Sun, describing
a short orbit from several hours~\cite{Ginzburg2020} to a few minutes~\cite{Burdge2022}. Millisecond pulsars achieve fast high spin rates when accreting material from a normal (i.e. non-degenerate) binary companion~\cite{Manchester2017}, typically  $\sim 0.1 \, M_{\odot}$, and have been discussed as a possible source of the gamma-ray excess observed in the region surrounding the Galactic Center.

As emphasized in the Introduction above, the standard mechanism of neutron star cooling predicts stellar surface temperatures of $\it{T_s}$  $< 10^4$ K for stars ages $\tau > 10^7$ years~\cite{Kouvaris:2007ay}. On the other hand, the ultraviolet thermal emissions detected from MSP older than $10^7$ years, imply an increase in their surface temperatures to $\it{T_s}$ $ \sim 10^5 K$ or more ~\cite{Deller2008, Andreas2010}. Black Widow pulsars, on the other hand, are thought to be much older, and could feature higher $\it{T_s}$ values (now, just upper limits are set) \cite{Gentile2014}. In this work, we address this problem in order to shed some light on the effects of late heating mechanism(s) in older NSs.

In order to address this problem, we perform in the following a comprehensive study of heating and cooling effects of a large set of observed  pulsars.
More precisely, we explore additional heating mechanisms of these stars that, if present, would cause NSs to reach higher  equilibrium temperatures involving heating and emission processes, depending of course on the nature and intensity of the additional heating sources.
Among the heating mechanisms we consider dark matter (DM) annihilation inside the cores, --- causing late-time heating ---, magnetic field decay and the effects of slowdown in the pulsar rotation which drives the matter out of beta equilibrium, resulting in an imbalance in the chemical potentials, the so-called late-time rotochemical heating \cite{Andreas2010}. 

We adopt in the following an ideal  gas model for the pulsar composition consisting of a mixsture of neutrons, protons, and leptons. In view of the lack of knowledge of the fundamental nature of DM, we consider the case where the capture rate has maximum efficiency ($f = 1$) and consider fermion DM particles falling into the widely known acronym of WIMP category, {\it Weakly Interacting Massive Particles}, that is, particles with mass and interaction strength around the electroweak  scale, based on the assumption they have a high accumulation rate in compact stars~\cite{Bell2018}. Among the best-known candidates are the fourth generation of neutrinos, i.e.,  a {\it heavy neutrino} (Dirac or Majorana), as well as supersymmetric SUSY particles, more specifically a {\it neutralino}, --- a linear superposition of the fermion partners of the neutral electroweak gauge bosons and of the neutral Higgs bosons (Higgsinos) ---,  and the {\it sneutrino}, the scalar super-symmetric partner
of the neutrino, since it is electrically neutral and weakly-interacting. We may consider also alternative proposals beyond the SM as {\it Universal Extra Dimension models} and {\it Little Higgs theories}, among others~\cite{WIMPs}. The efficiency $f$ in turn depends on the relationship between the scattering DM-neutron cross section and a saturation cross section\footnote{The saturation cross section depends on the neutron star geometric cross section $\sigma_{0}= \pi (m_n/M_\star)R_\star^{2} $. We assume one single scattering for GeV $\leq  M_\chi \leq 10^6 GeV$ and that all kinetic energy from the halo and gravitational energy binds the DM to the NS. Hence, we have $\sigma_{s}=\sigma_{0}$. For DM mass outside this range, the saturation cross section scales differently (see \cite{Masha2017,Camargo:2019wou}).}, above which all incident DM is captured.

\section{Modeling the EoS}
A minimal model of a NS, complying with all the recent NICER structural constraints \cite{NICER} is required to proceed with the calculations. We have study four EoS; two of them were constructed by using the model of Negele $\&$ Vautherin \cite{NV} for the crust and their cores were fitted by a polytropic EoS. In order to find physical constraints, they must be comply with the NS maximum mass and the measured radius by NICER\cite{NICER} simultaneously. A polytropic EoS is used due to its simplicity for implementation. A large set of hadronic EoSs can be fitted by three polytropic models with a fixed crust\cite{Read}.  In addition, we also make the predictions for the EoS AP4(npl)\cite{Lattimer:2000nx} used in the rotochemical mechanism and MPA1(npl)\cite{Lattimer:2000nx}. Note that MPA1 automatically respects both NICER and the maximum observed masses, as will be show in the section describing the results. Both AP4 and MPA1 EoS were employed in tabulated form.
The polytropic EoS is giving by\cite{Zdunik:2005kh}
\bea
&P(n) &= K n^{\Gamma} \, ; \\
&\epsilon (n)& = \frac{P}{\Gamma-1} + m_0 n \, .  
\eea
In this expression, $P$ represents the pressure (MeV/fm$^3$), $\epsilon$ the energy density (MeV/fm$^3$), $n$ is the baryon number (1/fm$^3$) and $m_0$ (with $c =1$), the adjusted baryon mass (MeV). Moreover, $\Gamma$ is the adiabatic index and the second coefficient, $K$, is often called the pressure coefficient. 
The fitted values are listed in Table \ref{table1}.
\begin{table}[htbp]
\tbl{Parameter values for the sets of polytropic EoS.}
{\begin{tabular}{@{}cccc@{}} \toprule
Set & K [MeV/$n^{\Gamma-1}$] & $\Gamma$ &
$m_{0} [MeV]$  \\ \colrule
1&  1842.64	& 3.18   & 945.66  \\
		\hline
		2 &	632.20    & 2.8   & 945.30  \\ \botrule
\end{tabular} \label{table1}}
\end{table}
Set 2 in Table \ref{table1} has a similar M-R relation for the MPA1 EoS.
\section{Galactic halo DM in NSs}
The capture of galactic DM particles by a NS occurs mainly through their scattering with Standard Model (SM) particles.
The high density and strong gravity of a NS may ``compensate'' the feeble DM interactions and enhance their capture rate~\cite{Keung2020}. Naively estimating, an old NS can be heated to equilibrium temperatures within the near-infrared band of the blackbody spectrum, from $\it{T_s}$  $\sim 100$ K to $\sim $ 2000 K.
In this work we consider the scattering between DM particles with neutrons only.
As we have seen, the threshold capture cross section, which is equivalent to the saturation cross section for DM$_{\chi}$-neutron scattering is of the order of $3\times 10^{-45} cm^2$, --- for NS values of ${R} = 12 km$ and {\it{M}} = $1.4 M\odot$ ---, and may heat an old NS up to an equilibrium temperature of the order $T_{s} \sim 10^3 \, K$ for DM-masses values ranging between $GeV$ and $PeV$. For masses lighter than $\sim \ 1 \ GeV$, the capture rate is suppressed by Pauli blocking effects while for mass values greater than $\sim \ 1 \ PeV$, it is necessary to consider multiple scattering to decelerate the halo particles~\cite{Keung2020}.
For DM particles with mass around $GeV - TeV$, self-dispersion in galactic halos exhibits quantum-mechanical resonances, analogous to Sommerfeld enhancement for annihilation~\cite{Tulin2012, Keung2020}.

The growing number of high quality observations
of thermal radiation from ancient NSs are still limited to a range of temperatures and ages. In the near future, infrared telescopes will measure temperatures down to a $\sim$ few thousand Kelvin, making it possible to investigate stellar heating through DM capture processes. In addition to the dark kinetic heating and dark annihilation heating which can be elastic or inelastic scattering off SM particles, the {\it decay} of DM  may affect thermal radiation processes, a hypothesis still far from a proper understanding\cite{Andreas2010}.

 As stated, the capture of DM particles by the gravitational potential of a NS occurs when the energy transferred by them as a result of their collision with neutrons reduces their initial kinetic energy in the galactic halo to a threshold below escape. Since we seek to determine the maximum value of the surface temperature of the NSs, as stated before we set the capture efficiency value to $f=1$. For the details on the $\sigma_{\chi n}$ and $\sigma_{sat}$ cross-sections see refs. ~\cite{Masha2017, Bell2018}.
\subsection{DM galactic halo profile}
For the description of the galactic DM halo we adopt Einasto's
profile~\cite{Einasto1965}, characterized by a power-law logarithmic slope,
\begin{equation}
\rho(r) =  \rho_s   \exp\Bigl\{ -\frac{2}{\alpha} \Bigl[ \Bigl( \frac{r}{r_s} \Bigr)^{\alpha} - 1 \Bigr] \Bigr\} .
\end{equation}
In this expression, the parameter $\alpha$, $r_s$ and $\rho_s$ are adjusted to reproduce DM in galaxies and its rotational curve. In this work we adopted the following values for the parameters of Einasto profile, normalized to local DM density of $0.3 \ GeV/cm^3$: $\rho_s = 0.054 \ GeV/cm^3$, $\alpha = 0.17$, and $r_s = 20.3 \ Kpc$ \cite{Navarro:2008kc}.
\subsection{DM capture rate: Model G1}
The capture rate $F$ of DM particles that remain trapped inside the NSs is given by Model G1~\cite{Goldman1989}:
\begin{equation}
\quad F = \frac{\pi b^2_{max} v_{\chi} \rho_{\chi} f}{m_{\chi}}; \quad \mbox{with} \quad
f = min(\sigma_{\chi n}/\sigma_{sat},1)  \in [0,1] \, , 
\end{equation}
where $b_{max}$ is the maximum impact parameter given by 
\begin{equation}
b_{max}=\Biggl(\frac{2GM_\star R_\star}{v_{\chi}^2}\Biggr)^{1/2}\Biggl(1 - \frac{2GM\star}{c^2R_\star}\Biggr)^{-1/2} \, .
\end{equation}
Additionally $\rho_{\chi}$ is the DM local density, $m_\chi$ is the DM mass, and  $f$ is the fraction of dark particles passing through the star that become trapped in its interior, and depends on the DM-nucleon (DM-N) cross section, $\sigma_{\chi n}$;
$\sigma_{sat}$ is the saturation cross-section, for which all transiting DM particles are captured by the stellar matter of the NS interior. This model assumes a Schwarzschild metric for a test particle grazing the star surface.

The power released by this model is
\be
W= \frac{\pi b_{max}^2 v_\chi \rho_{\text{dm}} f E_{s}^{R}}{m_{\chi}} \, ,
\ee
where 
\begin{equation}
E_{s}^{R} \simeq m_\chi(\gamma_{esc}-1+\chi) \, , \quad \mbox{with} \quad \gamma_{esc}=(1- (v_{esc}^2/c^2))^{-1/2} \, ,
\end{equation}
is the total energy which can be deposited by DM at the core of a NS and $\chi$ is the fraction of DM-annihilation. The thermalization occurs in $\sim$ {\it{Myr}}, corresponding to mature NSs. For dark kinetic heating $\chi=0$, for both dark kinetic and annihilation or decay heating, $\chi$=1.  Also, for the dark mass range considered here-$ GeV \lesssim m_\chi \lesssim PeV$, --- DM needs just one scattering to be captured by the NS. Considering that a DM-N scattering event depletes the DM kinetic energy, in the rest frame of a NS, by the quantity
\begin{equation}
\Delta E_s= \frac{m_nm_\chi^2 {\gamma_{esc}}^2 v_{esc}^2 (1-cos(\theta_c) )}{m_n^2 +m_\chi^2+ 2 \gamma_{esc}*m_\chi*m_n} \, ,
\end{equation}
where $\theta_c$ is the scattering angle in the center-of-mass frame and $v_{esc}=\sqrt{2GM_\star/R_\star}$ is the escape velocity at the interaction point. A way to see the PeV limit of DM mass is by comparing the initial kinetic energy in the halo  to the rest blue-shifted NS frame, $\frac{\gamma_{esc} m_\chi v_\chi^2}{2}$. Taking $m\chi >> m_n$ and equating these expressions, we find that 
\begin{equation} 
m_\chi \simeq 2\gamma_{esc} (v_{esc}^2/v_\chi^2)m_n \simeq \mbox{PeV} \, ,
\end{equation}
i.e the maximum mass for a DM particle to be captured by one scatter.
 Yet, the condition\cite{Goldman1989} of gravitation to overcome the Pauli blocking and accumulate to form a black hole is m$_\chi$= $10^8$ GeV in $10^8$ {\it{yr}}. The calculations were performed using the parameters {\it{R${_\star}$}}=10 {\it{km}}, {\it{ M${_\star}$}}=1.4$M_\odot$, v=250 {\it{km/s}} and $\rho_{\text{dm}}= 0.375 $  {\it{GeV/cm$^3$}} (for calculations with other parameters sets, see ref. \cite{Goldman1989}).

\subsection{DM capture rate: Model G2}
This model takes into account relativistic effects, and its capture rate is  \cite{Kouvaris:2010vv},
\begin{equation}
F=\frac{8}{3} \pi^2 \frac{\rho_{\text{dm}}}{m_\chi} \left ( \frac{3}{2 \pi
	v_{\chi}^2} \right )^{3/2} \Biggl( \frac{GM_{\star}R_{\star}}{1-\frac{2GM_{\star}}{c^2R_{\star}}}
v_{\chi}^2f \Biggr) \, , \label{kouvrel}
\end{equation}
where $\rho_{\text{dm}}$ is the DM density at the NS location, $m_\chi$ is the DM mass, $R_\star$ and $M_\star$ are the radius and mass of the NS, $v_\chi$ is the average velocity of WIMPs far enough from the star and $f$ is the parameter associated with the trapping efficiency of DM inside the star. In this expression, $f$ is equal to one for elastic or inelastic collisions, in this case only if the inelastic cross section is greater than $\sim 10^{-45}~\text{cm}^2$, otherwise, $f=\sigma_N/(10^{-45}~\text{cm}^2)$ for
$\sigma_N<10^{-45}~\text{cm}^2$. Eq (\ref{kouvrel}) was derived on basis of ref.~\cite{press}.
The power released by the annihilation of DM inside the star is given by\cite{Kouvaris:2007ay}
\be
W(t)=F m_\chi \,\text{tanh}^2\Bigl(\frac{t+c}{\tau}\Bigr),
\label{power1}
\ee
where $c$ is a small quantity that depends on the initial conditions and can be neglected; the capture time is $t$ and $\tau=1/\sqrt{FC_A}\ $. The constant $C_A=<\sigma_{\chi}v>/V$ is the thermally averaged annihilation cross section over the effective volume within which the annihilation takes place. For $t>>\tau$, which is the case we are analyzing for old neutron stars,  $tanh^2(\frac{t+c}{\tau})$ saturates to 1. Therefore,
\begin{equation}
W \simeq F \, m_\chi \, \chi,
\label{power2}
\end{equation}
where $\chi$ is the fraction of heat energy. The coefficient $\chi$ depends on the DM candidate and its self-annihilation channels. Here, we assume that $\chi=1$, i.e., that all energy is deposited locally.
It is important to point out that, in order to carry out these mechanisms that DM particles trapped in a NS first collide with the neutrons, and at least after one collision (for DM masses between 1 TeV and 1 PeV) they achieve thermal equilibrium within the star. This is because, once the typical life span of the NS considered in this study have $t > 10^6 \ $ {\it{yr}}, the thermalization have already been reached after one collision and no transient contribution has to be considered. It is well-known that after an age $t \approx 10^{5-6} \, yr$, the main contribution to the NS cooling are due to photons emission from the surface of the star~\cite{Page2015}, reflecting in turn its internal temperature.

The normal heat sources can compete with the heat from DM annihilation\cite{Kouvaris:2007ay} for late times, when the star temperature is low enough to cool by photons emission only. In this stage, the power released through DM annihilation, Eq (\ref{power2}), equals the thermal energy loss rate $L_{\gamma}$. It is worth to note that once the equilibrium is reached, the temperature remains constant, i.e., it does not drop further. In 
ref.~\cite{Kouvaris:2010vv}, the author characterized Eq  (\ref{kouvrel}) as ``relativistic'' due to the difference of the classic formulation by the factor $1/(1-(2GM_{\star}/c^2 R_{\star})$, which is $\sim 1.52$ for the values of R$_\star$ and M$_{\star}$ assumed in this article. 
\section{The rotochemical mechanism}
For a NS in chemical equilibrium described by an ideal gas composed of neutrons ({\it{n}}), protons ({\it{p}}), and leptons(electrons and muons), we have $\eta_{npl}\equiv \mu_n -\mu_p -\mu_l=0$. Nevertheless, if we consider that the star is slowing down, its centrifugal force is reduced, the central density increases and the chemical potential becomes $\eta_{npl}~\ne~0$. As the chemical composition changes, the star will then reach a new chemical equilibrium due to beta decays, releasing energy both in the form of neutrinos and heat, later radiated as photons. It is important to mention that this holds for the non-superfluid case, since nucleons ``locked'' in
paired states would not participate in the reactions.
The chemical imbalance evolution is given by \cite{Fernandez:2005cg}
\be \label{roto}
\dot{\eta}_{\text{npl}}=-A(\eta_{\text{npl}},T)-R_{\text{npl}}\Omega\dot{\Omega} \, , 
\ee
where the scalar quantity $R_\text{npl}$ quantifies the departure from equilibrium due to the change in the angular velocity and the function {\it{A}} quantifies the effect of reactions toward restoring chemical equilibrium. The luminosity created by this mechanism is $L= \Gamma {\eta}_{\text{npl}}$, where $\Gamma= \Gamma_{n \rightarrow pl\bar{\nu}}- \Gamma_{pl \rightarrow n\nu} $. The rate reactions are integrated over the core and correspond to transformation of neutrons to protons through a direct or modified URCA process. The second term of the rate represents the inverse reaction. For proper evaluation of this mechanism it is necessary to solve both Eqs (\ref{roto}) and (\ref{cool}) (see next) simultaneously. It is important to note that rotational slowdown evolution is assumed to be due to magnetic dipole radiation, without any magnetic field decay. Accordingly, we will have one initial value for the temperature and another one for the magnetic field. In the following we adopt the AP4 EoS, which allows only the modified URCA processes, with the initial internal temperature $T = 10^{8} \, K$  with ${M}=1.4 M_{\odot}$ and magnetic field $B = 10^{8} \, G$. We do not solve these equations, but instead adopt the internal temperature as a function of time from ref.~\cite{Andreas2010} in  Eqs (\ref{TS}) and (\ref{TSAP}) (see below). 

In the following we adopt the canonical NS mass value, $1.4 M_{\odot}$, and the radius of $12 \, km$, although other values are clearly possible \cite{NOS}.
We observed that the authors of ref. \cite{bellt}
considered not the rest NS mass in the EoS, but its effective mass, since at the core the strong interactions become relevant. Also, DM is accelerated to quasi-relativistic velocities upon infall into a NS. This requires a momentum dependence of the neutron form factors, since it is not seen as a point-particle anymore, but as a composite one with quarks and gluons. However, since the transfer momentum is not large enough, deep inelastic scattering does not make a crucial change in the total cross section. Regarding the pseudo-scalar operator, the authors found nevertheless a suppression of three orders of magnitude compared to free fermion gas without momentum dependence when considering the QMC EoS with $1.9 \, M_{\odot}$ as an example. In order to compare their results with ours we  arbitrarily
choose the value of the cutoff scale ($\Lambda$). The upper limit from collider physics is  $\Lambda \leq  M_{\chi}/2\pi$  
for the contact operators approach~\cite{Bai:2010hh}. Accordingly, the DM-N cross-section can result in a drastically reduction in the surface temperature of the star once determined the cutoff scale of the theory. For instance, a DM with mass of $500 \, GeV$ result in $\Lambda \sim \, 80 \, GeV$. However, we can not guarantee the validity of such a constraint for the interior of a neutron star.
\section{Magnetic Field Decay: a simple estimate}
A possible source of heating inside NSs is the decay of the magnetic field, dissipating energy into heat. The intensity of the magnetic field is rarely high enough for dramatic effects, but for completeness, an estimation is presented below.
We used a simple approximation for the decay of a magnetic field\cite{Andreas2010} and surface photon emission luminosity for $\tau$ $>$ $10^6$ {\it{yr}}. Assuming a star with radius $R_{\star}$ and magnetic energy {\it{E$_{B}$}} $\sim$ $(4\pi {\it {R{_\star}}}^3/3)<B^2>/8\pi$ in a time scale $t$, the luminosity estimate is simply
\be
L\sim \frac{E_{B}}{t}\sim \frac{4\pi R{_\star}^3}{3}\frac{<B^2>}{8\pi t}. \label{L}
\ee
To obtain the emissivity of magnetic field decay we must divide the luminosity by the star volume. Note that in the expression above we have an explicit dependence on time as in the rotochemical mechanism (see below). A magnetic field of $10^{10}$ {\it{G}}, as measured for BWs systems \cite{Hui:2019pin}, has been used. Internal magnetic fields can have much higher values and could change this estimate if these are the fields that decay and dissipate.
\section{Luminosities and Cooling Mechanisms}
	 The luminosity is a fundamental quantity to evaluate the cooling of compact stars. The heating loss rate from the surface of the star (in its local reference frame) is
	\be \label{lumi}
	L_{\gamma}=4 \pi
	R^2 \sigma T^4_{*} , 
	\ee
	where $\sigma$ is the Stefan-Boltzmann constant and {\it{$T_{*}$}} is the surface temperature. However, we must convert the surface temperature to the reference frame at infinity, namely
	\be \label{TS}
	T_{s}= T_{*}\sqrt{1 - \frac{2GM_\star}{c^2R_\star}}.
	\ee
	 It is worth to note that for old NSs some works address late time values, and not the time evolution along the star's life. In this way, the heating mechanism with a specific luminosity is given by Eq (\ref{lumi}) (considering the star as a blackbody emitter). Thus, it is possible to estimate the surface temperature from the  reference frame at infinity.
	Moreover, the internal uniform temperature (core) is always higher than the crust. The relation of the surface temperature to the internal core temperature can be approximated by \cite{Gundmundsson1,Gundmundsson2,Page:2004fy}
	\be \label{TSAP}
	T_{s}=8.7 \times 10^5  \left(
	\frac{g_s}{10^{14}\text{cm/s}^2} \right)^{1/4} \left(
	\frac{T}{10^8 \text{K}} \right)^{0.55} {\it{\text{K}}} ,  \ee
	where $T$ is the core temperature of the star, assumed to be an uniform quantity,  and $g_s=GM_\star/{R_\star}^2$ is the surface gravity.
	We can re-write Eq (\ref{lumi}) in terms of its internal core temperature as
	\be \label{15}
	L_{\gamma}=4 \pi R^2
	\sigma (8.7 \times 10^5 K )^4 \left( \frac{g_s}{10^{14}
		\text{cm/s}^2} \right) \left( \frac{T}{10^8 \ }
	\right)^{2.2} \text{{\it{erg $s^{-1}$}}}.
	\ee
	Here we are considering only NSs that do not achieve extremely high densities, so that only the modified URCA process (a bystander neutron assisting the direct URCA process) can occur. Under these conditions, the direct URCA processes
	$n~\rightarrow~p~+~e~+~\bar{\nu}$, $p~+e~\rightarrow~n~+~\nu$ are kinematically forbidden. During the first cooling stage, the star loses energy due to neutrino emission, by converting protons and electron to neutrons and vice-versa. The emissivity for modified URCA process is \cite{Shapiro:1983du}
	\be
	\epsilon_{\nu}=1.2 \times 10^4 \left( \frac{\rho}{\rho_0}
	\right)^{2/3} \left( \frac{T}{10^7 \text{K}}\right)^8  \text{erg} ~ {\it{ \text{cm}^{-3} \text{s}^{-1}}}.
	\ee
	The direct URCA process, usually depending on the EoS, is subject to the density condition  $\sim$ 4$\rho_0$, where $\rho_0$ is the standard nuclear matter density. In the following, we assume\cite{Yakovlev1995}
 $\rho_0= 2.8 \times 10^{14}{\it{g \ cm^{-3}}} $ .
	The surface temperature in Eq. (\ref{15}) is related to the interior temperature $T_{\rm int}$ using the relation (\ref{TSAP}).
	The differential equation that governs the cooling of a compact star is given as 
	\bea \label{cool}
	\frac{dT_{\text{int}}}{dt}&=&\frac{-L_{\nu}-L_{\gamma}+L_{\text{h}}}{V
		c_V} ,
	\eea
	where $c_V$ is the specific heat for a gas of non-interacting fermions, $L_{\nu}$, $L_{\gamma}$ and $L_{\text{h}}$ are the luminosity of neutrinos, photons and heating mechanism(s). In this expression $c_V$ is given by
	\be
	c_V=\frac{k_B^2T}{3 \hbar^3 c} \sum_i
	p_F^i\sqrt{m_i^2c^2+(p_F^i)^2},
	\ee
	where the sum runs over neutrons, protons and electrons and with the Fermi momenta (for neutral matter) in weak equilibrium given as  \cite{Shapiro:1983du}
	\begin{eqnarray} & p_F^n &=340\left(\frac{\rho}{\rho_0}\right)^{1/3} ~\text{{\it{MeV}}}, \\
		& p_F^p & =p_F^e=60 \left(\frac{\rho}{\rho_0}\right)^{2/3}
~\text{{\it{MeV}}}.	\end{eqnarray}
	After the integration of  Eq (\ref{cool}), we need to account the effects of the envelope combined with Eqs  (\ref{TS}) and (\ref{TSAP}). Notice that the gas of non-interacting fermions does not behave as in the rotochemical mechanism, since it takes into account the effective nucleon mass and changes in the EoS chemical potential due to rotation effects. As we said before, our approach assumes an ideal gas mixture for nucleons and leptons.
  \section{Application to young and intermediate age NSs}
  The surface temperatures attributed to the thermal emission of NSs have been studied over the years. The first task is to disentangle the thermal emission from non-thermal processes (magnetospheric, winds) which also contribute to the total emission. For young objects the typical expected surface temperature $T_{s}$ peaks around $\sim 1 \, keV$. The book of ref. \cite{book1} contains a summary of these observations for objects which are younger than $\sim 10^{6} \ {\it{yr}}$. There is a trend for a decrease of the temperature with increasing age, and a visible drop between $\leq 10^{6} \ {\it{yr}}$ when the photon emission begins to dominate.

  The physics leading to the NSs cooling outcomes is affected by several uncertainties, both in the microphysical and astrophysical domains. Among the most important ones we point out the existence of superfluid gaps between nucleons in the outer layers, the very composition of superdense matter (not only because of the plausible presence of deconfined quarks, but also in the ordinary nuclear phase even if the former are absent), and the fallback of stellar matter producing a different composition of the atmosphere, among many others. High magnetic fields substantially modifies the cooling, opening or suppressing new channels, which may be the cause of the high temperatures inferred for magnetars, a subgroup believed to be highly magnetized and very young.

  As a result, it is quite hopeless to attempt an evaluation of DM effects for the main set of young NSs. In fact, there are hints of {\it accelerated  cooling}, in which magnetic fields or direct URCA neutrino processes may be involved (PSR J1740+1000, Vela, see \cite{pothe}). The spread of the observed temperatures is substantial \cite{book1} but the trend of cooling for the first $\sim 10^{6} \ {\it{yr}}$ is clear.
  \section{Results: pplication to old NSs and the case of the Black Widow pulsars}

   The number of older NSs measured is scarce. This is unfortunate, because it is around $10^{7} \, \ {\it{yr}}$ or so when the late heating mechanisms would be more relevant. In the following a few results are shown in the following figures. Fig. \ref{fig1} presents the thermal flux measured from Earth which will be discussed below. Fig. \ref{fig2} shows the mass-radius relation from solutions of Tolman-Oppenheimer-Volkoff equations\cite{tov} using as input the following EoSs: AP4, MPA1, set1 and set2. The mass-radius relation will fix the ratio $\rho/\rho_{0}$ for each maximum mass and its respective radius. In Fig. \ref{fig3}, we show the capture rate for DM models G1 and G2  corresponding to the MPA1 EoS which are compatible
results with NICER and Pulsar’s masses measurements. The surface temperature (blue) for young neutron stars were obtained from ref. \cite{Beznogov:2014yia} while for the middle NSs, the green upper limits and black error bars are from ref. \cite{Yanagi:2019vrr}. In Fig. \ref{fig4}
we show the behaviour of the effective surface temperature in terms of time for the ratio  $\frac{\rho}{\rho_0} = 1.61$.
  For the determination of surface temperatures of old objects, is it therefore important to address the heating mechanisms. One potentially interesting case is the group of {\it Black Widow} binary systems, in which the pulsar thought to be recycled is now seen to be ablating its companion, at least completely. Evolutionary calculations suggest
  ages of $\sim several \, Gyr$ for this group, potentially probing the extreme of the age distribution.
Unusual physical elements such as back-illumination and ablation winds are important to explain these systems in the orbital period-companion mass plane \cite{roberts,linares}. The evolutionary tracks calculated in ref. \cite{Benvenuto2012} have been used to infer an age for the systems shown in Figs. \ref{fig2}, \ref{fig3} and \ref{fig4}. A somewhat different scenario~\cite{chengetal} also gives comparable ages, in the ballpark of several $Gyr$, which are very old and suitable to test the internal heating hypothesis.

  X-ray observations~\cite{Gentile2014} constitute a source of information about surface temperatures $T_{s}$. The separation of the non-thermal components presented in that work rendered temperatures $T_{s}$ in the ballpark of $\sim 100 ~\it{eV}$. If correct, these results suggest that NSs would not cool substantially after their first million years. However, the extracted values correspond to small emission surfaces, and have been linked to ``hot spots''. We have used the black body fit from ref.~\cite{Gentile2014} and these surfaces temperatures (orange data) should be considered as upper limits in Figs. \ref{fig4},\ref{fig5} and \ref{fig6}. It remains to be seen if a true $T_{s}$, not just upper limits, can be obtained for the surface temperature with improved observations and analysis.

\section{Model's limitations vs surface temperature observations}
\begin{figure}[htbp]
	\centering
			\includegraphics[scale=0.75]{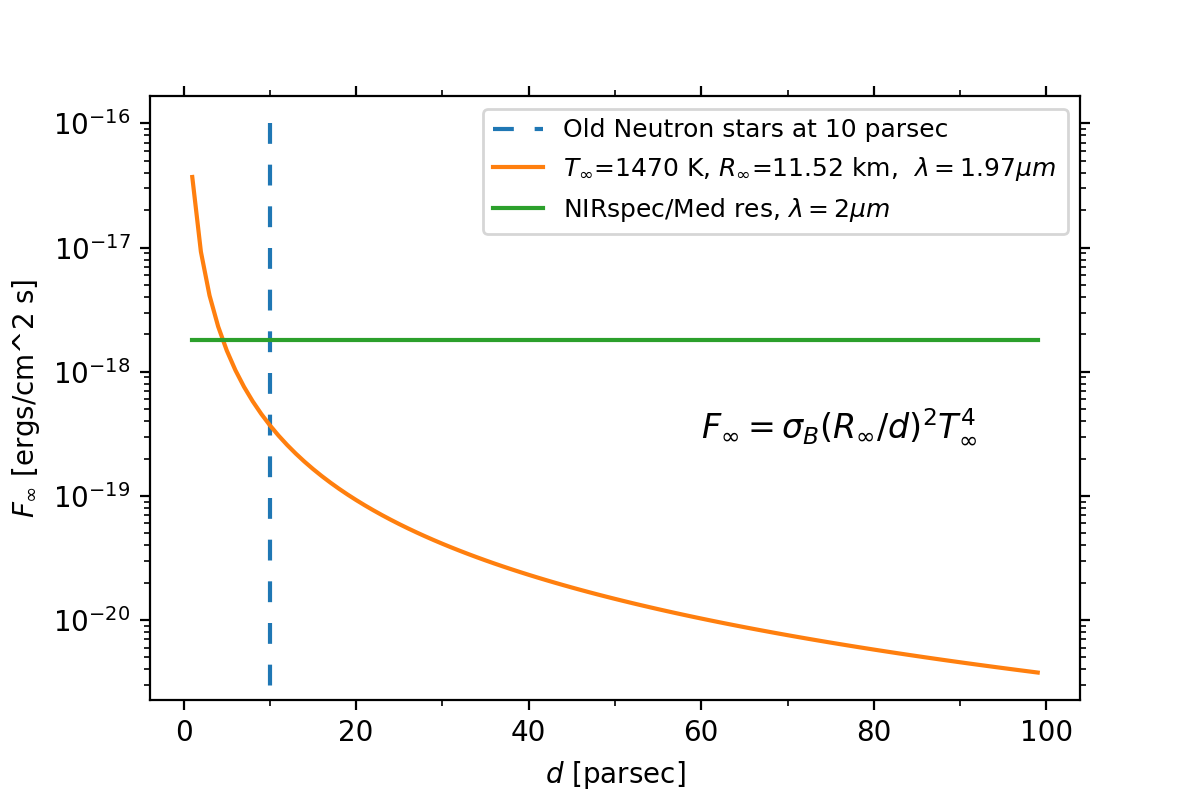}
					\caption{Thermal flux in terms of distance from Earth. The old neutron stars at $10 \, pc$ were considered in ref.~\cite{1993ApJ...403..690B}.}%
					\label{fig1}%
				\end{figure}

Our simulations for the evolution of the temperature in terms of star's age could be considered as a toy model to guide the discussion, since it does not use the atmosphere models, and we considered the EoS of the particles as a free fermion gas. As expected, the effects of DM heating do not effect the surface temperature of young NSs since it is not temperature-depend and the standard picture renders much higher values for $T_{s}$. Moreover, for older NSs, the predictions for the surface temperature are substantially below the observed in their upper limits. The situation for dark kinetic heating is not improved if we consider the highest surface temperature together with NICER constraints (Fig. \ref{fig1}), which results in $T_{s}=1470 \, K$ for capture efficiency $f = 1$ (optimal case) and the thermal flux resolution-brightness of a point source detected at $10 \, \sigma$ in a $10,000 s$ integration using the James Webb space telescope (JWST)\cite{jameswebbsite}. Accordingly, 
the prediction is almost one order below the sensitivity of the JWST. The thermal flux could be achieved increasing the exposure time. Nevertheless, the uncertainty on capture efficiency depends on the cross section DM-neutron, which is model dependent, and the cross section threshold. This model dependency is explored in refs.~\cite{Camargo:2019wou,Bell2018,Raj:2017wrv}.    
Moreover, recent studies\cite{Gonzalez-Jimenez:2014iia,Hamaguchi:2019oev} favors the rotochemical heating model as the one which describes the surface temperature of old NSs (note that BWs are not addressed in that work, and they will not provide firm new information until their true $T_{s}$ could be determined).  As pointed out in ref.~\cite{Gonzalez-Jimenez:2014iia} the rotochemical heating is best suitable for MSP PSRJ0437-4715, i.e., an old milisecond pulsar.  	
  \section{Results: comparison of heating scenarios and the corresponding data}
Tables \ref{tabela2}-\ref{table3} and Figs. \ref{fig5}-\ref{fig10} show our comparison results of heating scenarios and the corresponding data. Figs. {\ref{fig5}, similarly to Fig. {\ref{fig4}
show the effective surface temperature in terms of time for the ratio  $\frac{\rho}{\rho_0} = 1.61$. Fig. {\ref{fig6}
shows the surface temperature evolution plotted as a function of time for $\frac{\rho}{\rho_0} = 3.52$. The magenta line for no DM falls before the $10^8 \, yr$. We conclude that the larger the ratio, the further to the right all curves move. Fig. \ref{fig7} presents an estimate of DM density for BWs and the PSR J1740+1000 using the Einasto's halo profile. The results show a difference of three orders of magnitude between predictions and observation.
Fig. \ref{fig8} shows the surface temperature $T_s$ for DM capture taking into account the limits of radii and masses found in ref.~\cite{Ozel:2016oaf} for  $\rho_{\text{dm}}=0.3 {\it{GeV/cm^3}}$ and {\it{V$_\chi$=230 km/s}} based on the model from ref.~\cite{Bell2018}. For the local DM density, the highest surface temperature is about 1760 {\it{K}}. Fig. \ref{fig9} presents similar calculations as in the previous figure; taking the best fit for DM density the surface temperature corresponds to $1900 \, K$. Notice that both cases correspond to $R_{\star} = 10 \, km$ and $M = 2 \, M_{\odot}$. In Fig. \ref{fig10}
similar calculations are shown. This figure uses the limits imposed by NICER and the local DM density. The highest surface temperature ($1470 \, K$  occurs when $R_{\star} = 11.52 \, km$ and $M_{\star} = 1.49 \, M_{\odot}$.

  The heating mechanism(s) do not seriously affect the younger NSs, i.e., with age less than $10^{7} \, yr$. However, the surface temperatures for mature NSs are compatible only with the magnetic field decay and rotochemical heating but incompatible with the upper limits of Black Widows if their true $T_{s}$ happen to be near the former.
   The set of parameters used in the models requires some comments. As we have shown, there are two ways to obtain the surface temperatures: in the first, we use only the luminosity for late stages, and in the second, the differential equation for the cooling star is used. We have compared both approaches, which lead to almost the same result. However, ignoring the differential equation, the stages of cooling and heating can be overlapped. For the rotochemical model (for the parameter set used by the authors of ref.~\cite{Andreas2010}) this is not a problem, since the heating mechanism begins to be important after {10$^6${\it{yr}}} but for DM models the situation is different. We also searched for the maximum luminosity (heating process) that starts after  $\approx$  $10^6$ {\it{yr}} for model G1 and magnetic field decay approximation.
   \begin{table}[htbp]
\tbl{Results for models G1 and G2 with the following parameters: {\it{R$_\star$=12 km, v=230 km/s, M$_\star$=1.4M$_\odot$ and $\rho_{\text{dm}}$=0.3 GeV/$cm^3$}}.}
{\begin{tabular}{@{}cc@{}} \toprule
Model &  T$_{s}$ [ {\it{K}}]  \\ \colrule
	G1 ($\gamma_{esc}=1.23$)  kinetic &    1390          \\
   		\hline
   		G1 ($\gamma_{esc}=1.23$) kinetic+annihilation & 2100            \\
   		\hline
   		G2        &     2165      \\
   		\hline
   		G2 ({\it{R=12.2 km, M=2.2 M$_\odot$}}, MPA1 $\sim$ $4\rho_0$)       & 2650 \\
   		\hline
   		J0023+0923 (BWs, central value)   &  2088 $\times$ $10^3$ \\ \botrule
\end{tabular} \label{tabela2}}
\end{table}
  
   Table \ref{tabela2} shows the predictions of surface temperatures for models G1 and G2 compared with the J0023+0923 (BW) data.
      As expected, depending on the  aforementioned parameters, the results change significantly. Comparing the value of surface temperature\cite{Gentile2014} from  BWs J0023+0923 with the DM model, a difference of three orders of magnitude is obtained. For the case of BWs, even using the favorable G2(A) model (MPA1 values for radius and mass), the models are too cool and would not explain high surface temperatures. This illustrates the importance of new studies of the Black Widows temperatures. Some authors argue that near the galactic center or in galaxy clusters these results could change, since the $\rho_{\text{dm}}$ can be higher.
   Dividing the surface temperature of G2(A) model by 1.52 (to transform to its classical version) we obtain $\sim 1400 \, K$. This MPA1 value is quite close to the corresponding one in the model G1(K). The two models have different derivations, but comparing the case G1(K+A) with G2(A), we found that their values are very close and far from the observed $T_{s}$.

  \begin{table}[htbp]
\tbl{Results from differential equation of cooling at time $10^9 \, yr$  for surface temperature in Kelvin. The last row is the ratio between the surface temperature  of J0023+0923 and MPA1
predictions with  $R = 12.2 \, km$ and $M = 2.2 M_\odot$.}
{\begin{tabular}{@{}cccccccc@{}} \toprule
$\frac{\rho}{\rho_0}$ & G2(A) & G1(K)  & G1(K+A)  & MFD  & Lmax & Rotochemical \\ \colrule
   		1.61	& 1969 & 1264 & 1915  & 50504  & 124972 &123262   \\
			\hline
			2.47	& $\sim$ & $\sim$ & $\sim$ &$\sim$  &$\sim$  & 123262  \\
			\hline
			3.52 (MPA1)	&2232 & 1697 &2263   & 47373  & 115774 &  123262\\
			\hline
			J0023+0923/MPA1	& 935  & 1230 & 922 & 44  & 18    & 17  \\ \botrule
\end{tabular} \label{table3}}
\end{table}
In Table \ref{table3} we show the prediction of surface temperature for all models; G2(A) stands for DM model G2 considering only annihilation; DM model G1(K) takes only the dark kinetic heat into account; G1(K+A) considers both dark kinetic and annihilation heating; MFD stands for magnetic field decay and $L_{max}$ for the maximum luminosity allowed after the time $10^6 \, yr$.
The ratio J0023+0923/MPA1 predicts a difference of three orders of magnitude between the pulsar surface temperature and the corresponding results for the MPA1 EoS in the DM models, rotochemical mechanism and magnetic field decay. The maximum luminosity was intended to extrapolate the DM mechanism. We may find, for instance, the amount of $\rho_{\text{dm}}$ needed for a maximum luminosity. Further, the increase of $\frac{\rho}{\rho_0}$ shifts the slope for a slightly longer time. 

For a NS located at some distance from Earth, we used the Einasto's profile getting $\sim$ three orders of difference between the DM models and the observed surface temperature of NSs and BWs. Also, we get a ``coincident' result for rotochemical, magnetic field decay and maximum luminosity. All these describe a new measure of the PSR-B0950+08\cite{Hai-tao:2022izf}. Thus, a more robust mechanism inspired in magnetic field decay and DM (annihilation and kinetic heating) could be appropriate to describe isolated NS, since the rotochemical mechanism is enough for that.
In order to compare with a third model, we used Eq (2.10) from ref.~\cite{Bell2018}. The surface temperature for the aforementioned parameters is $\sim$ 1394 {\it{K}} since it considers only dark kinetic heating. Also, this model takes into account the dispersion velocity of NS and the velocity of DM. The same assumptions as before (radius, mass and v$_\chi$) were used. In addition, we evaluate Eq. (2.10) for constraints with the NICER \cite{NICER} and ref.~\cite{Ozel:2016oaf}. These 3D plots show {\it{T$_{s}$}} for a variety of radii and masses for each DM density: local density  of DM and the best fit for the Milky Way\cite{Pato:2015dua} which are 0.3 {\it {GeV/cm$^3$}} and {\it{0.42 GeV/cm$^3$}} both for $v = 230 \, km/s$.
\begin{figure}[H]
	\centering
		\centering
		\includegraphics[scale=0.675]{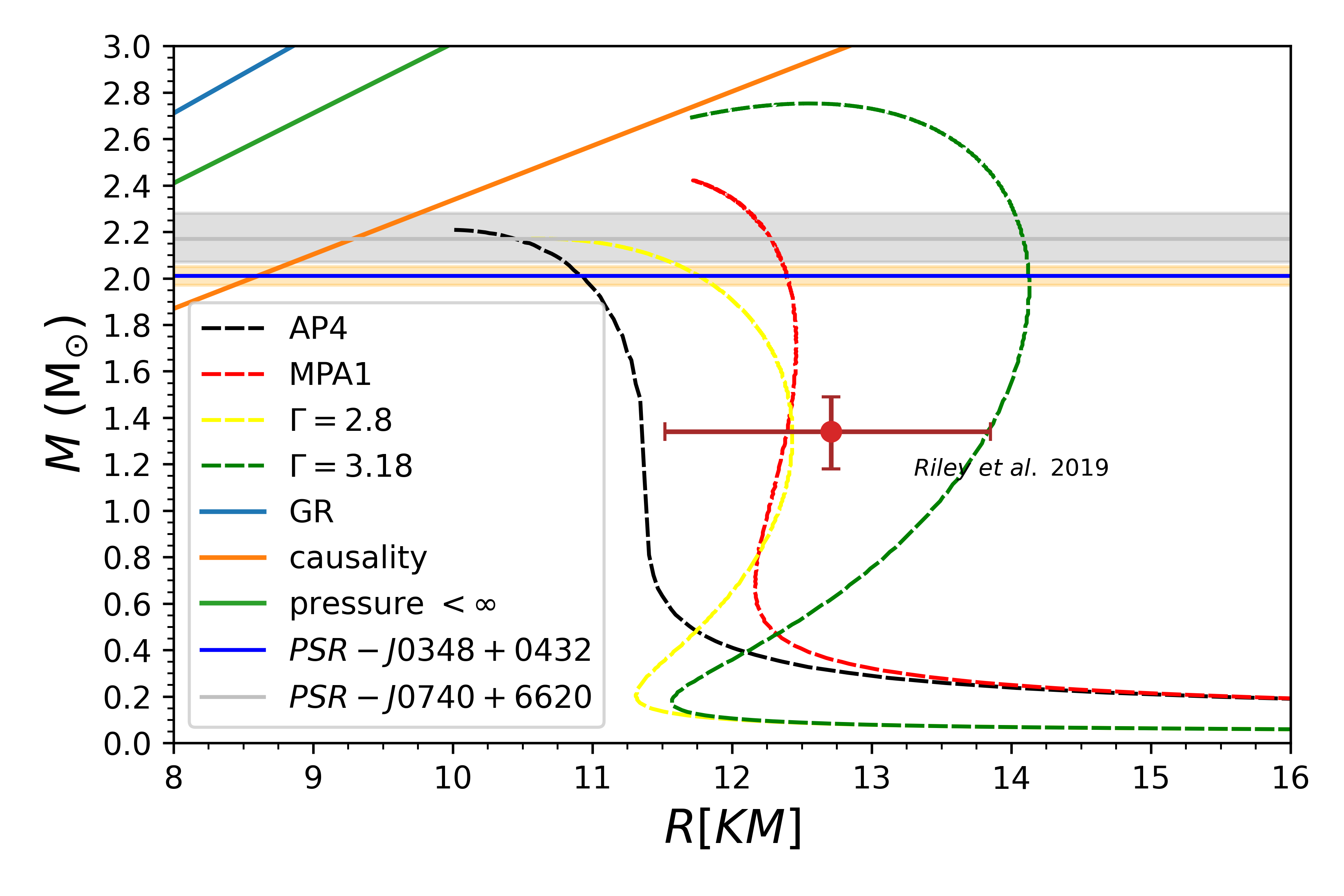}
\caption{Mass-Radius relations for different EoSs. The limits of pressure, causality and General Relativity parameters are from ref.~\cite{Lattimer:2006xb}.}
\label{fig2}
\end{figure}	
	\begin{figure}[H]
		\centering
		\includegraphics[scale=0.7]{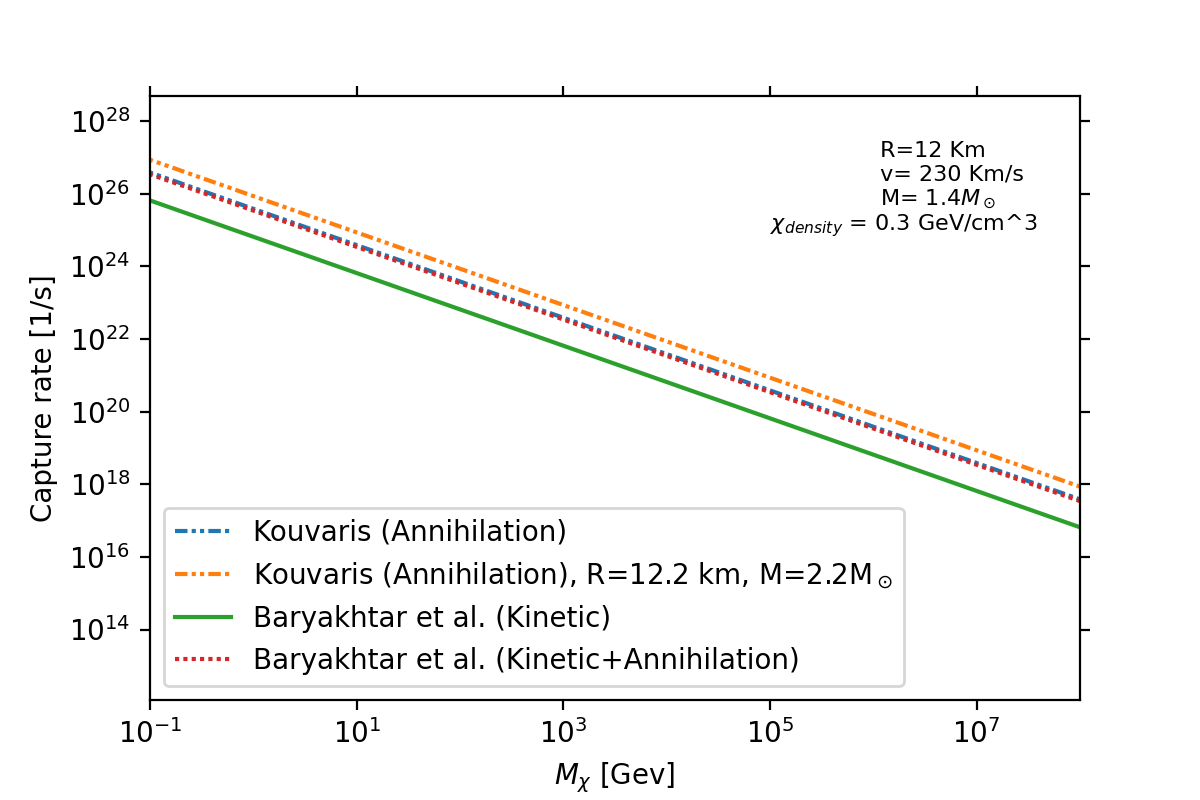}
\caption{The capture rate for DM models G1 and G2 is depicted. MPA1 EoS shows compatible results with NICER and Pulsar's masses measurements. The AP4 EoS, however, agrees only with the measured masses. Our two sets are in accordance with NICER and pulsar's masses. The capture rate was calculated for 2 cases. For the first case, we have $R_{\star} = 12.0 km$, $v_\chi = 230 \, km/s$, $\rho_{\text{dm}} = 0.3 \, GeV/cm^{3}$ and $M = 1.4M_{\odot}$, while for the second, $R_{\star} = 12.2 \, km$, $v_\chi = 230 \, km/s$, $\rho_{\text{dm}} = 0.3 \, GeV/cm^{3}$ and $M = 2.2 \, M_{\odot}$. In this figure, {\it Kouvaris} stands for model G2 while  {\it Baryakhtar et al.} refers to model G1.   }
\label{fig3}%
\end{figure}
\begin{figure}[H]
		\centering
		\includegraphics[scale=0.75]{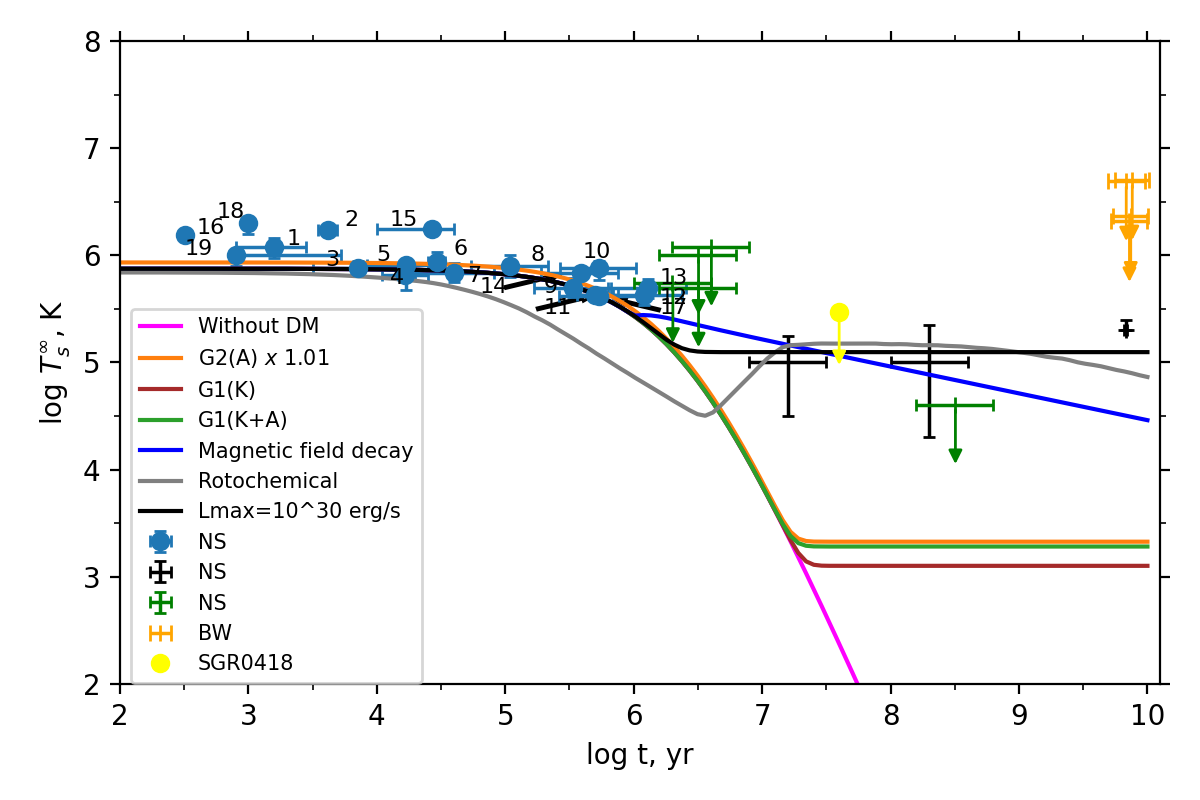}
	\caption{Effective surface temperature in terms of time for the ratio  $\frac{\rho}{\rho_0} = 1.61$}
	\label{fig4}
\end{figure}
\begin{figure}[H]
		\centering
		\includegraphics[scale=0.75]{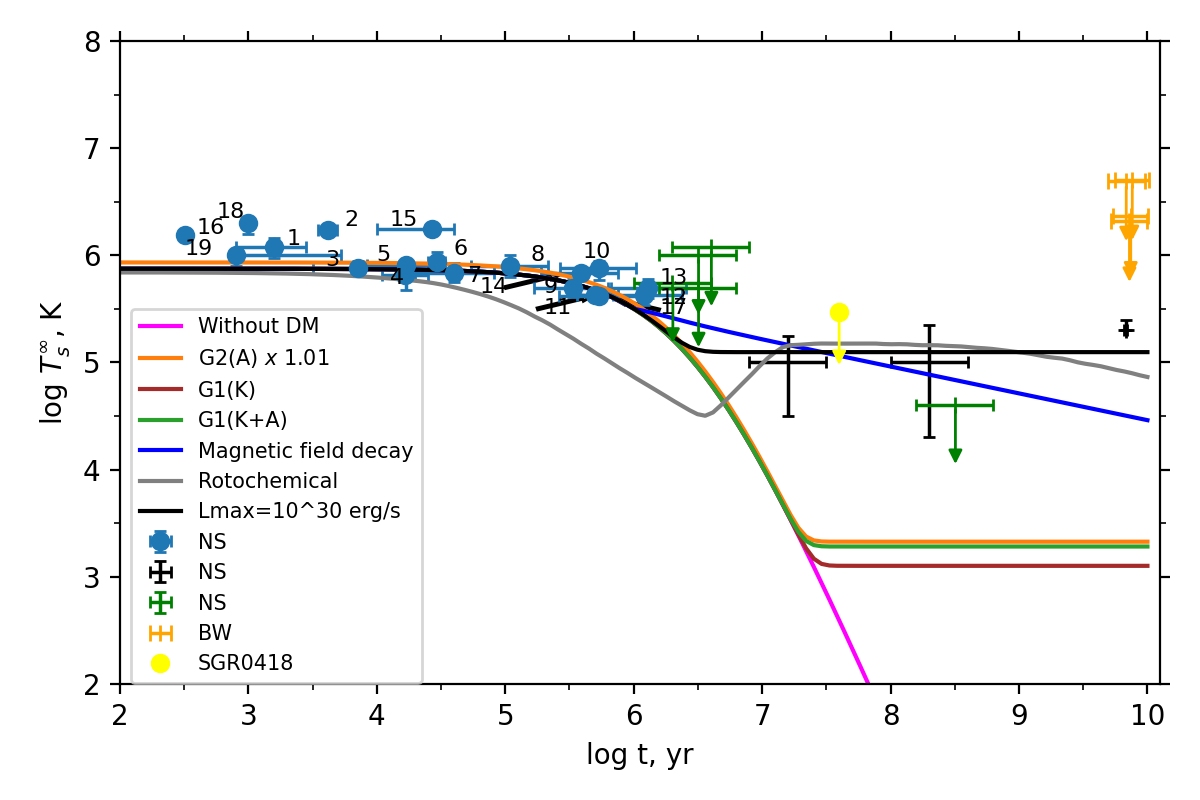}
	\caption{Similar calculations as the previous figure but for $\frac{\rho}{\rho_0} = 2.47$. Notice the slightly deviation to right of all curves, except the one associate with the rotochemical mechanism .}%
\label{fig5}%
\end{figure}
\begin{figure}[H]
	\centering
		\includegraphics[scale=0.7]{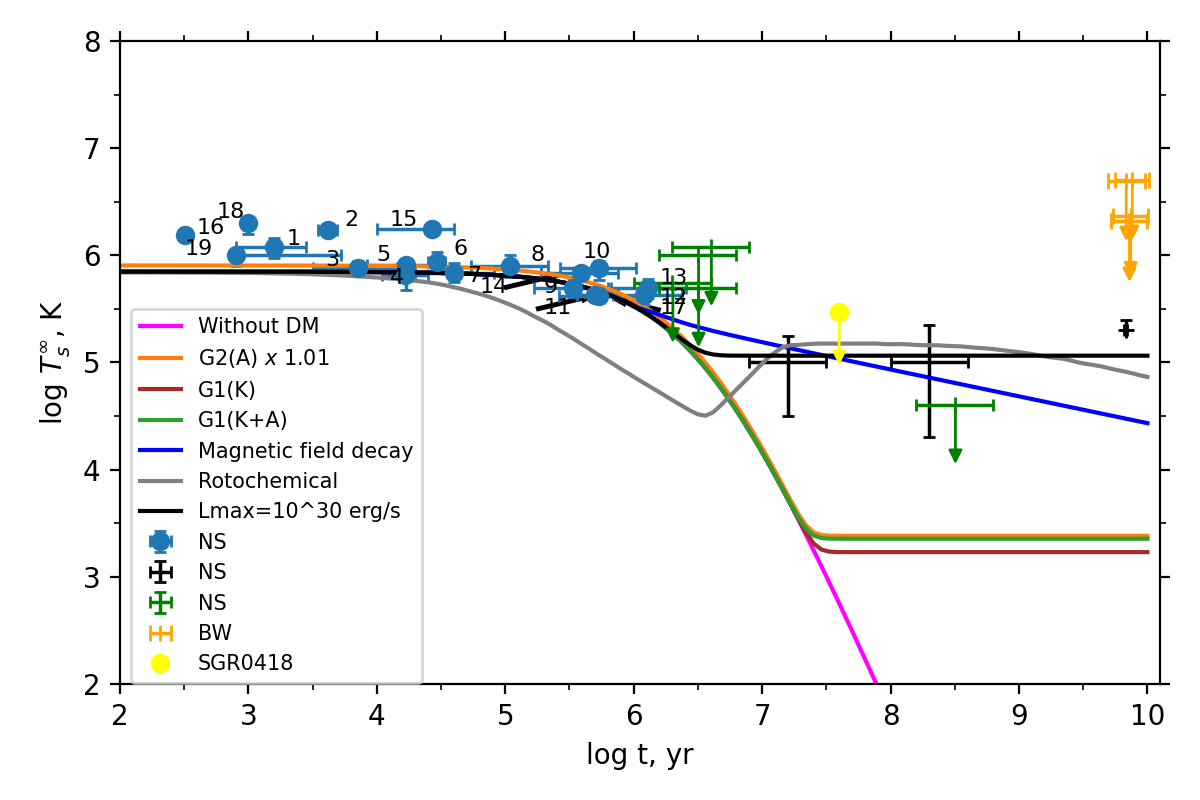}
		\caption{The surface temperature evolution is plotted as a function of time for $\frac{\rho}{\rho_0} = 3.52$. The magenta line for no DM falls before the $10^8 \, yr$.}
		\label{fig6} 
	\end{figure}
	\begin{figure}[H]
		\centering
		\includegraphics[scale=0.75]{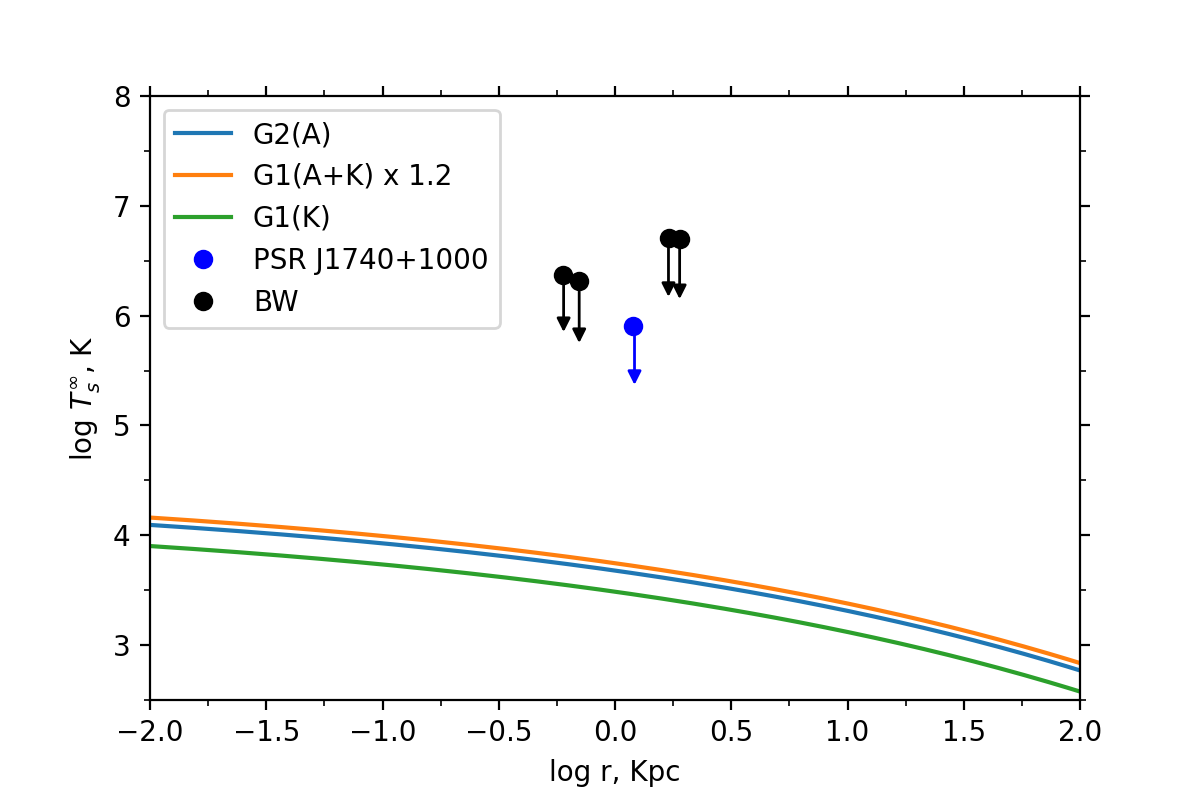}
		\caption{Surface temperature in terms of distance (r) using the Einasto's halo profile. For close distance, the dark matter halo profile results in higher density.}%
		\label{fig7}
	\end{figure}
	\begin{figure}[H]
		\centering
		\includegraphics[scale=0.8]{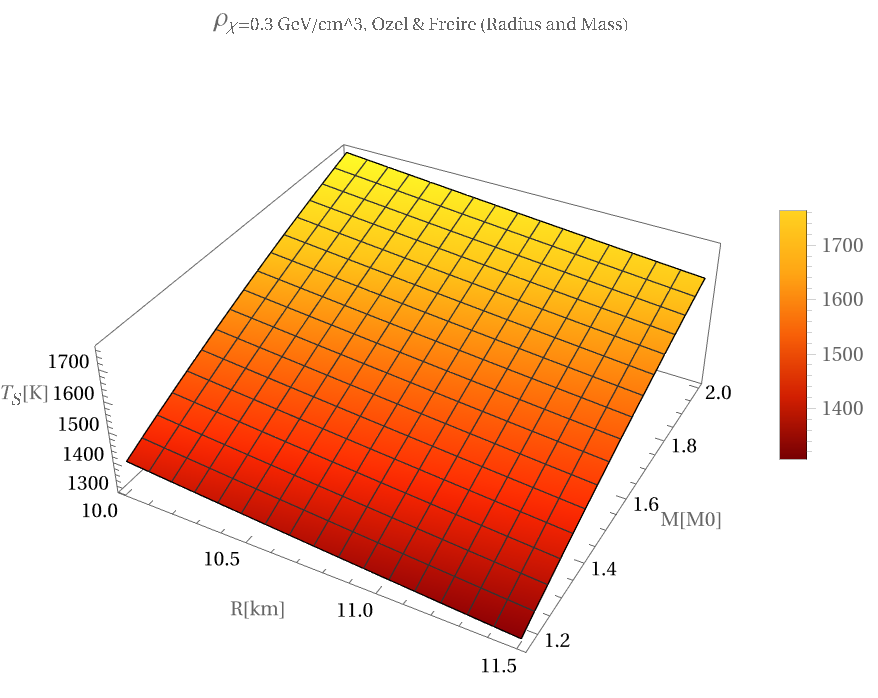}
		\caption{Surface temperature $T_s$ for DM capture taking into account the limits of radii and masses found in ref.~\cite{Ozel:2016oaf} for  $\rho_{\text{dm}}=0.3 {\it{GeV/cm^3}}$ and {\it{V$_\chi$=230 km/s}} based on the model from ref.~\cite{Bell2018}. }
		\label{fig8}
	\end{figure}
	\begin{figure}[H]
		\centering
		\includegraphics[scale=0.8]{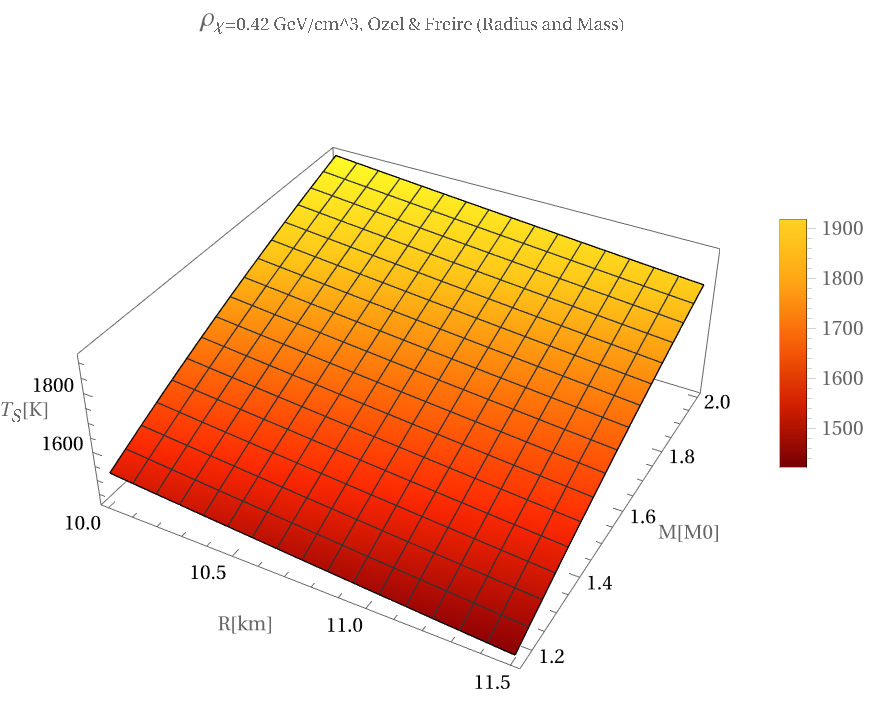}
		\caption{Similar calculations as in the previous figure; taking the best fit for DM density the surface temperature is $1900 \, K$. }
		\label{fig9}
	\end{figure}
\begin{figure}[H]
	\centering
\includegraphics[scale=0.8]{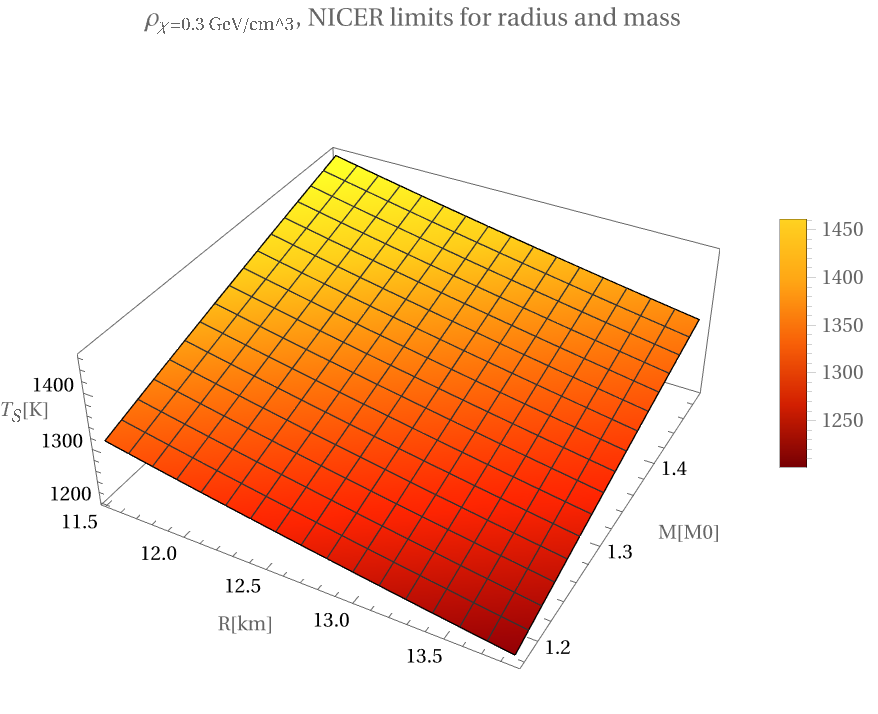}
\caption{Similar calculations as in the previous figure. This figure uses the limits imposed by NICER and the local DM density. The highest surface temperature ($1470 \, K$  occurs when $R_{\star} = 11.52 \, km$ and $M_{\star} = 1.49 \, M_{\odot}$. }
	\label{fig10}%
\end{figure}

\section{Conclusions}
In this contribution, three DM models and one magnetic field decay heating mechanism were studied in order to obtain predictions for old NSs including BWs. Our results indicate that for young NSs these mechanisms are not strictly necessary.
However, at later times, we found that rotochemical process are the best candidate to explain the surface temperatures of mature NSs, since the magnetic field decay in a crude approximation shows a constant decrease in $T_{s}$, while the luminosity is consistent with a $\sim$ constant surface temperature. In order to find the maximum luminosity, we set a magnetic field which start to decay considerably after $10^{6} \, yr$. Furthermore, we have considered several stars' radii and masses consistent with observations. These results are shown in 3D plots, which comprise not just one surface temperature, but a wide set of them. Einastos's DM halo profiles were employed to determine the dark matter density at the position of the NSs/BWs. The difference between the model
predictions and actual surface temperature is large, up to three orders of magnitude. The most
dramatic differences are found for the case of BWs: all heating mechanism would clearly fail to
describe their surface temperatures even if they happen to be well below the current upper limits, provided their ages exceed $\sim \, a \, few \, Gyr$ .

Since BWs feature the largest known NS masses \cite{linares,jfinal}, the latest report	being $2.35 \pm 0.17 \, M_{\odot}$ by Romani et al. \cite{romani} in a study of PSR J0952-0607, it is tempting to suggest a connection between the former and high temperatures at old ages. However, this naive idea would require elaboration. The presence of exotics (i.e. quarks, condensates, etc.) are generally found to {\it accelerate} cooling, not delay it. In addition, none of the ``spiders'' with determined temperatures have their masses measured, and therefore it is not certain that they fall in the supermassive category.
\section*{\Large{ Acknowledgments}}
JEH wish to acknowledge the financial support of the Fapesp Agency (S\~ao Paulo) through the
grant 2020/08518-2 and the CNPq (Federal Government, Brazil) for the award of a Research
Fellowship. P.O.H. acknowledges financial support from DGAPA-PAPIIT (IN100421). F.K. acknowledges the financial support of the CNPq (Federal Government, Brazil).
We would like to thank the referee for comments that improve the article.

\end{document}